\documentclass{webofc}

\usepackage[varg]{txfonts}   
\usepackage{hyperref}
\usepackage{url}
\usepackage{cleveref}
\usepackage{braket}

\newcommand{\madgraph}{\textsc{MadGraph5\_aMC@NLO}}
\newcommand{\mg}{\textsc{MGaMC}}
\newcommand{\cudacpp}{\texttt{CUDACPP}}

\usepackage{algorithm}
\usepackage{algpseudocode}

\crefname{algocf}{alg.}{algs.}
\Crefname{algocf}{Algorithm}{Algorithms}

\hypersetup{colorlinks=true,citecolor=blue,urlcolor=blue,linkcolor=blue}
\begin{document}
\title{Hardware acceleration for next-to-leading order event\\generation within \madgraph{}}
%
%

\author{\firstname{Zenny} \lastname{Wettersten}\inst{1,2}\fnsep\thanks{\email{zenny.wettersten@cern.ch}} \and
        \firstname{Olivier} \lastname{Mattelaer}\inst{3} \and
        \firstname{Stefan} \lastname{Roiser}\inst{1} \and
        \firstname{Andrea} \lastname{Valassi}\inst{1} \and
        \firstname{Marco} \lastname{Zaro}\inst{4}
}

\institute{European Organization for Nuclear Research (CERN)
\and
           Technical University of Vienna (TU Wien) 
\and
           University of Louvain (UCLouvain)
\and
            University of Milan (UniMi)
          }

\abstract{As the quality of experimental measurements increases, so does the need for Monte Carlo-generated simulated events — both with respect to the total amount and to their precision. In perturbative methods, this involves the evaluation of higher order corrections to the leading order (LO) scattering amplitudes, including real emissions and loop corrections. Although experimental uncertainties today are larger than those of simulations, at the High Luminosity LHC experimental precision is expected to be above the theoretical one for events generated below next-to-leading order (NLO) precision. As forecasted hardware resources will not meet CPU requirements for these simulation needs, speeding up NLO event generation is a necessity.

In recent years, collaborators across Europe and the United States have been working on CPU vectorisation of LO event generation within the \madgraph{} framework, as well as porting it to GPUs, to major success. Recently, development has also started on vectorising NLO event generation. Due to the more complicated nature of NLO amplitudes this development faces several difficulties not accounted for in the LO development, but it shows promise. Here, we present these issues as well as the current status of our event-parallel NLO implementation.}
\maketitle
\section{Introduction}
\label{intro}
Over the course of the upcoming High-Luminosity LHC (HL-LHC) era, the integrated luminosity is expected to rise by an order of magnitude when compared to the end of Run-3 \citep{Schmidt_2016}. The measurements made at the LHC need to be statistically compared with Monte-Carlo (MC) generated simulated events, and the predicted precision of HL-LHC experiments necessitate not only an order of magnitude increase in the number of generated events, but also of their precision \citep{HSFPhysicsEventGeneratorWG:2020gxw}. In perturbative quantum field theory (QFT), this entails moving from predictions made at leading order (LO) to ones at next-to-leading order (NLO).

In response to this issue, several MC event generators have made efforts to port event generation to data-parallel architectures, such as GPUs and vectorised CPUs \citep{HSFPhysicsEventGeneratorWG:2020gxw,Yazgan:2783019,Valassi:2021ljk,Valassi:2022dkc,Valassi:2023yud,Hageboeck:2023blb,valassi_2024_14940556,Bothmann:2023gew}. For the general-purpose event generator \madgraph{} (\mg{}), this has been done at the level of LO \textit{tree-level} scattering amplitudes, where significant speed-up has been found. Continuing this effort, we have started work on parallelising NLO amplitudes, where many challenges --- both old and new --- are found. Nevertheless, we have developed a multithreaded prototype for event-parallel fixed order calculations in \mg{}, which is currently in the process of being validated. Efforts to port these developments to event generation and to inject tree-level amplitudes with vector instructions are planned.

Due to the developmental status of the parallel fixed order implementation, we here focus on the fundamental difficulties of NLO calculations. In \cref{sec:pqft,sec:evgen}, we detail the theoretical background of perturbative QFT and MC event generation respectively. \Cref{sec:paranlo} extensively details \mg{} NLO calculations, both in the upstream codebase and in our multithreaded prototype, and elaborates on the technical issues differentiating NLO from LO. Finally, we summarise the concerns and current status in \cref{sec:outlook}.

\section{Perturbation theory}
\label{sec:pqft}

There are many stages between a QFT described in terms of a Lagrangian and the expected experimental observations, but the first one --- the one describing the underlying fundamental physics --- lies in going from a Lagrangian (describing all of the possible interactions in a model) to a cross section (the probability of a particular interaction actually occurring). In \cref{sec:dsigma} we briefly discuss the formulation of a closed form solution for the cross section using perturbation theory, and in \cref{sec:nlocontr} we detail the additional difficulties entailed in beyond-leading order calculations. For a thorough derivation, we refer the reader to the plethora of QFT textbooks available, e.g. \citep{book:peskinschroeder}.

\subsection{Differential cross sections}
\label{sec:dsigma}

Particle physics typically concerns the evaluation of cross sections. A problem arises here, though, as there is generally no closed form solution for the cross section. The standard way to handle this problem is \textit{perturbative QFT}, where we Taylor expand the interactions about the non-interacting theory. Mathematically, we express this as
\begin{align} \label{eq:perturbation}
    \lim_{T\to\infty(1-i\varepsilon)} \text{exp} \left[ -i \int_{-T}^{T} dt \, H_I(t)     \right] \simeq 1 - i \int_{-T}^{T} dt \, H_I(t) + \frac{1}{2} \int_{-T}^{T} dt \int_{-T}^{T} dt' \, H_I(t) \, H_I(t') + \dots
\end{align}
Ignoring the leading identity, the first non-vanishing term is referred to as the \textit{leading order} contribution, the second the \textit{next-to-leading order}, etc.\footnote{The LO term is not necessarily tree-level. For simplicity, we assume that the LO contribution is tree-level and forego any discussion on loop-induced processes --- they would, for our purposes, be considered NLO.}

In order to describe amplitudes in a clear and explicit way, we use \textit{Feynman diagrams}. Feynman diagrams are representations of the different ways the particles in an interaction can be connected to each other; particles are shown as lines, and interactions as nodes. As an example, consider the process $e^+ e^- \to 2\gamma$, electron-positron annihilation producing two photons. In the standard model, the LO Feynman diagrams for this process are\footnote{These are technically two different diagrams under photon interchange, as photons are indistinguishable.}
\begin{align}\label{eq:born}
    \frac{d \sigma_{LO}}{d \Omega} \simeq \left|\raisebox{-18pt}{\includegraphics[width=0.2\textwidth]{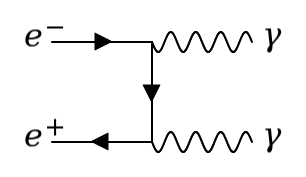}}\right|^2,
\end{align}
where $\sigma$ is the total cross section and $\Omega$ is the phase space --- the space of all momenta this interaction can take on. In \cref{eq:born}, the order of the diagram in powers of the electroweak coupling constant can immediately be read as the number of vertices --- nodes where different lines intersect --- and is equal to 2. As Feynman diagrams are, once evaluated, complex scalars, the differential cross section is given by the absolute square of their sum.

\subsection{Next-to-leading order corrections}
\label{sec:nlocontr}

Going on to the NLO contribution, we remind the reader that \cref{eq:perturbation} is a perturbative expansion in the interaction Hamiltonian. NLO thus necessitates an additional \textit{interaction}, i.e. an additional vertex. This can be thought of as appending an extra particle, and these contributions are referred to as \textit{real emissions}. Returning to $e^+e^-\to2\gamma$ as in \cref{eq:born}, some of the real emission diagrams are given by
\begin{align}\label{eq:real}
    \raisebox{-18pt}{\includegraphics[width=0.2\textwidth]{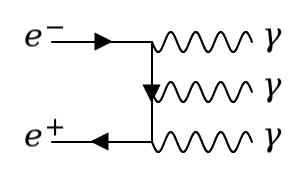}},
\end{align}
where as in \cref{eq:born} we must sum over all photon orderings. Furthermore, we must consider \textit{all} ways we can add an additional vertex, such as e.g. from a photon splitting into new fermions. Excluding other electroweak bosons, there are 54 real emission diagrams for this process. Contrast this to the two diagrams appearing at LO. 

Furthermore, we reiterate that $d\sigma$ is given by an absolute square, and at NLO we must also consider interference between the LO Born diagrams and \textit{loop amplitudes}. These arise when we add \textit{two} vertices to the LO diagrams with a completely internal particle, e.g.
\begin{align}\label{eq:loop}
     \raisebox{-18pt}{\includegraphics[width=0.26\textwidth]{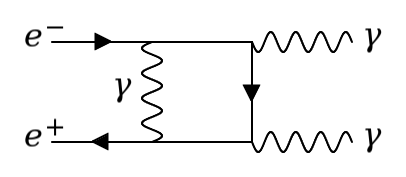}},
\end{align}
where the leptons first scatter against each other before annihilating. The term ``loop'' refers to the closed particle loop visible in the diagram; for \textit{tree-level} processes, such as those in \cref{eq:born,eq:real} the momenta of internal particles are uniquely defined by the external particles. For loop processes, however, the total momentum at each vertex depends on the unobservable internal propagator. Loop diagrams must thus be integrated over internal momenta. For $e^+e^-\to2\gamma$, again excluding other bosons, there are 19 loop diagrams.

The final complication of NLO amplitude computations arises during phase space integration --- $d\sigma_{NLO}$ has singularities across the phase space\footnote{Technically, $d\sigma_{LO}$ already has soft and collinear singularities, but they are by definition unobserved and thus excluded by the cuts enforced on the integral.}.
In fact, both the real and the loop contributions are singular in different parts of the phase space. However, since the cross section is physical, their \textit{sum} should be convergent across all of the phase space. This is proven in general in e.g. \citep{book:peskinschroeder}. For general evaluations, we employ different \textit{subtraction schemes} where we subtract zero in such a way that both the real and the loop contributions are locally convergent. Two standard subtraction schemes are FKS subtraction \citep{Frixione:1995ms,Frixione:1997np} and CS dipole subtraction \citep{Catani:1996vz}, the former of which is used in \mg{}.

\section{Parton-level event generation}
\label{sec:evgen}

\Cref{sec:pqft} provides an overview into the evaluation of observables, but misses one important aspect: QFTs are stochastic. What is observed at particle colliders will not be the expectation value of the theory but a statistical sample of events. To test a theory, we must compare experimental samples with predicted ones. The method used to generate these samples is called \textit{Monte-Carlo event generation} due to its basis in Monte-Carlo (MC) integration.

\subsection{Monte-Carlo integration}

MC integration is a numerical integration method based on the notion that an integral is necessarily equal to the mean value of the integrand times the volume of the integrated measure,
\begin{align}
    I = \int dx \, f(x) = \lim_{N\to\infty} \frac{1}{N} \sum_{n=1}^N f(x_n),
\end{align}
assuming that $x_n$ are randomly sampled across the integrated measure. The reason to use MC methods in HEP is twofold: Firstly, MC integration accuracy is independent of the dimensionality of the integrand \citep{press2007numerical}; and secondly, importance sampling the integral measure is exactly the same thing as determining the shape of a probability distribution. Thus, we can use MC techniques to first determine the total cross section of a process and then sample the phase space with the same likelihood we expect to see in an experiment. This process of phase space sampling is known as \textit{unweighted event generation} and is the first stage in QFT simulations when going from a theory to a quantifiable experimental prediction.

\subsection{Data-parallel LO event generation}
\label{sec:cudacpp}

Many event generators are employed for prediction and analysis at the LHC experiments; for an overview of some of the primary ones, see e.g. \citep{Buckley:2011ms}. Our focus will be on \mg{} \citep{Alwall:2014hca,Frederix:2018nkq}, a hard event generator\footnote{By a ``hard event generator'' we mean a generator of events according to the parton-level hard scattering process. The term event generator is somewhat ambiguous, but generally refers to the production of showered events.} supporting generic QFTs expressed in the UFO format \citep{Degrande:2011ua,Darme:2023jdn} with LO and NLO routines generated by ALOHA \citep{deAquino:2011ub}. \mg{} event generation generates amplitude routines for relevant parton configurations and determines the different phase space distributions before sampling them for unweighted events.

At LO, each parton configuration is identical across phase space. Each event evaluates the exact same function (\cref{eq:born}) for many phase space points. Consequently, \textit{``event-level parallelism looks like an appropriate approach to try and exploit efficiently [GPU and vectorized CPU architectures]''} \citep{HSFPhysicsEventGeneratorWG:2020gxw}. This was the motivation behind development of \cudacpp{}, a plugin for \mg{} which generates event-level data-parallel amplitude routines and interfaces them with \mg{} to run event generation on GPUs and vectorised CPUs \citep{Valassi:2021ljk,Valassi:2022dkc,Valassi:2023yud,Hageboeck:2023blb}. In our effort to utilise hardware acceleration also for hard NLO event generation, we expect opportunities to recycle much of the work done for LO, as elaborated on in \cref{sec:paranlo,sec:outlook}.

\section{Event-level parallelism at NLO}
\label{sec:paranlo}

To increase the precision of theoretical predictions in perturbative QFT, we must evaluate additional terms in \cref{eq:perturbation}. Given that the \cudacpp{} plugin presented in \cref{sec:cudacpp} enables hardware accelerated event generation at LO, the obvious next step is to port also NLO event generation to accelerators; we justify our decision to maintain an event-level parallel structure at NLO in \cref{sec:bottlenecks}. We then continue with the novel complications that arise when moving onto NLO in \cref{sec:nlocomplications}, and finally describe our efforts in \cref{sec:paraalgo,sec:status}.

\subsection{Compute bottlenecks in \madgraph{}}
\label{sec:bottlenecks}

\begin{figure}
    \centering
\sidecaption
    \includegraphics[width=0.7\linewidth]{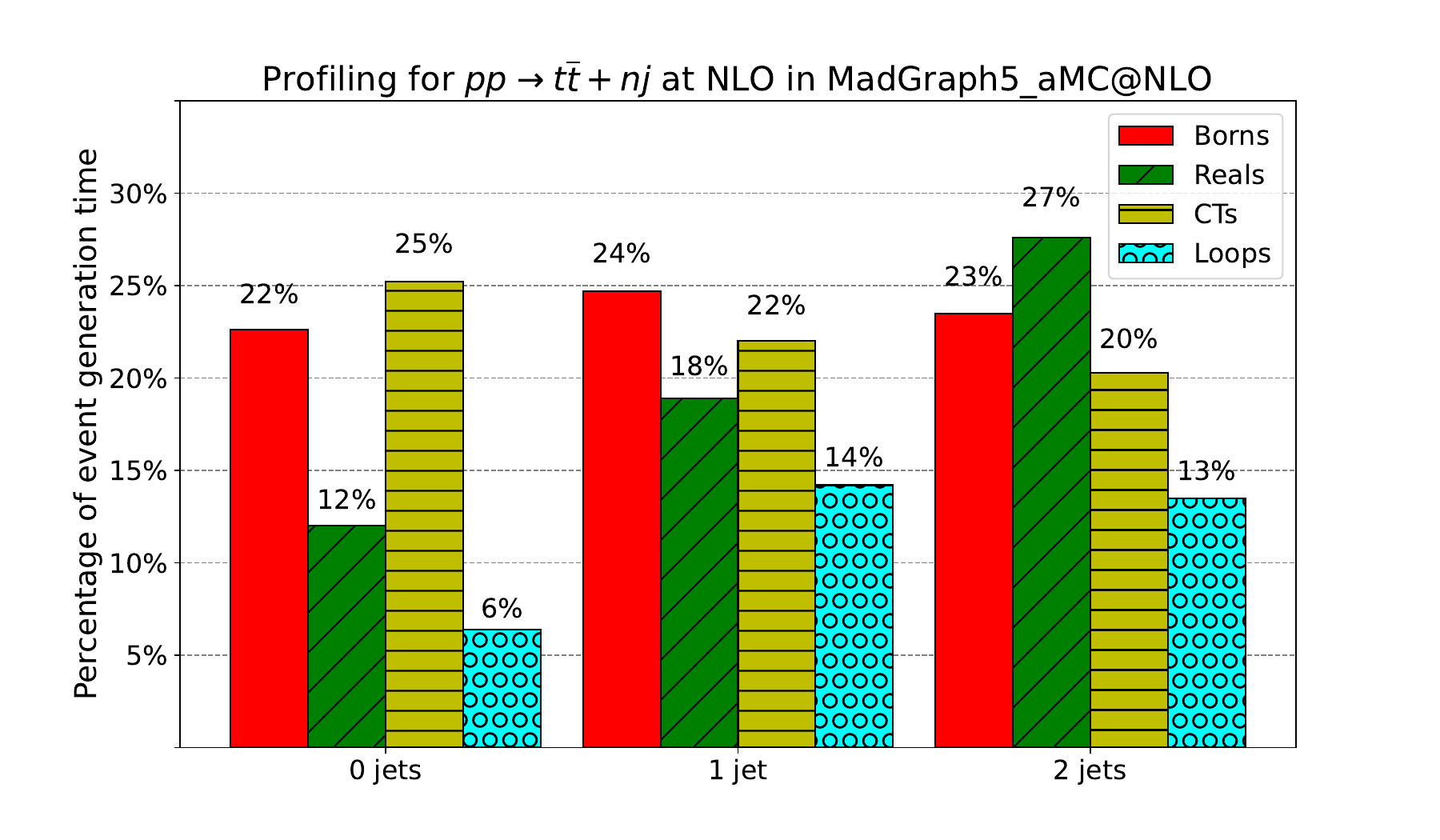}
    \caption{Runtime profile, generating 10 000 QCD NLO events in \mg{}.\\
    Borns refers to LO contributions; Reals (Loops) to real emissions (loop diagrams); and counter-terms (CTs) collectively to FKS subtraction terms and MC counter-events.\\
    Only amplitude evaluation runtimes shown.}
    \label{fig:nloprofile}
\end{figure}

In \cref{fig:nloprofile}, a runtime profile for pure-QCD NLO event generation in \mg{} is presented, considering the process $p\,p \to t \overline{t} \,+ nj$, $n=0,1,2$. Particularly, we only show the runtime fraction spent in Feynman diagrams --- summing, we find that scattering amplitudes make up $65-83\%$ of runtime for these processes. While not as dominant as at LO, amplitudes are still the primary bottleneck and just as at LO, they make up an increasing fraction of runtime as process complexity increases. Event-level parallelism still seems the appropriate approach.\footnote{As a sidenote, we mention that the large time spent in LO Born amplitudes is not representative of the runtime in LO event generation. Born amplitudes are evaluated at several different kinematics for the same event at NLO. Furthermore, these amplitudes are not recycled across evaluations, and are instead evaluated multiple times.}

\subsection{Complications in higher order evaluations}
\label{sec:nlocomplications}

While \cref{sec:bottlenecks} illustrates that parallelising scattering amplitudes still makes for the most obvious way to accelerate NLO event generation, certain fundamental differences between LO and NLO need to be addressed. There are a plethora of details that could be discussed, but we will limit ourselves to a couple key ones that need to be considered for hardware acceleration. In particular, these are phase space asymmetry and loop integrals.

An mentioned in \cref{sec:cudacpp}, amplitudes are identical across phase space for LO event generation. At NLO we need to consider the singular regions of the amplitudes as well as the local subtraction terms to ensure their convergence. Using FKS subtraction this is done by splitting the phase space into sectors with different kinematic parametrisations. As such, amplitudes are no longer identical; for SIMD and SIMT, parallelism should be across individual FKS sectors rather than the full phase space.

The second issue are the loop integrals. Firstly, they are computationally expensive\footnote{\Cref{fig:nloprofile} might give the impression that loops are not a problem, but the reason for this is that \mg{} generally uses an approximation of the full loop integral rather than explicit evaluation. For details, see section 2.4.3 of \citep{Alwall:2014hca}.}. This is not an issue in and of itself, but for the fact that \mg{} offloads loop evaluations to external libraries. Consequently, loop integration cannot be accelerated directly through \mg{}. Secondly, due to instabilities loop integration at times requires quadruple floating point precision (FP128). No major manufacturers have hardware-level support for FP128 instructions, neither for vectorised CPUs nor for GPUs.

One saving grace to read from \cref{fig:nloprofile} is that loop evaluations, at least for these pure QCD processes, are not a major bottleneck. A majority of the runtime is spent in LO Born and NLO real emission amplitudes and in CTs. It is important to note here is that \textit{all of these are tree-level scattering amplitudes.} That is, they all use the types of routines that have been vectorised for the \cudacpp{} plugin. Thus, we expect to be able to recycle the work on tree-level amplitudes from \cudacpp{} when parallelising NLO event generation.

\subsection{A data parallel algorithm}
\label{sec:paraalgo}

\begin{figure}
    \begin{minipage}{0.46\textwidth}
\begin{algorithm}[H]
    \centering
    \caption{Current event loop}\label{algo:mgevloop}
    \begin{algorithmic}[1]
        \State $FKS \gets \Call{setFKSsector}{\,}$
        \State $ p \gets \Call{genMomenta}{FKS} $
        \If{$\Call{passCuts\_nBody}{p}$}
            \State $\Call{M}{Born} \gets \Call{BornMatrix}{p}$
            \State $\Call{M}{Loop} \gets \Call{LoopMatrix}{p}$
        \EndIf
        \If{$\Call{passCuts\_Real}{p}$} 
            \State$ \Call{M}{Real} \gets \Call{RealMatrix}{p}$
        \EndIf
        \If{$\Call{passCuts\_Soft}{p}$}
            \State$ \Call{M}{Soft} \gets \Call{SoftMatrix}{p}$
        \EndIf
        \If{$\Call{passCuts\_Coll}{p}$}
            \State$ \Call{M}{Coll} \gets \Call{CollMatrix}{p}$
        \EndIf
    \end{algorithmic}
\end{algorithm}
\end{minipage}
\hfill
\begin{minipage}{0.46\textwidth}
\begin{algorithm}[H]
    \centering
    \caption{Branch-free event loop}\label{algo:paraevloop}
    \begin{algorithmic}[1]
            \State $FKS \gets \Call{setFKSsector}{\,}$
            \State $ p \gets \Call{genMomenta}{FKS} $
            \While{!\Call{passAnyCuts}{$p$}}
                \State $p \gets \Call{genMomenta}{FKS} $
            \EndWhile
            \State $\Call{M}{Born} \gets \Call{BornMatrix}{p}$
            \State $\Call{M}{Real} \gets \Call{RealMatrix}{p}$
            \State $\Call{M}{Soft} \gets \Call{SoftMatrix}{p}$
            \State $\Call{M}{Coll} \gets \Call{CollMatrix}{p}$
            \State $\Call{M}{Loop} \gets \Call{LoopMatrix}{p}$
            \For{$a \in [B,R,S,C]$}
            \If{$!\Call{passCuts}{a,p}$}
            \State $\Call{M}{a} \gets 0$
            \EndIf
            \EndFor
    \end{algorithmic}
\end{algorithm}
\end{minipage}
\end{figure}

Before describing our event-parallel NLO algorithm, we note that the codebases for LO and NLO in \mg{} are completely distinct, and the work on developing a multi-event interface at LO could not be reused. However, as mentioned in \cref{sec:nlocomplications}, the vectorised tree-level amplitudes in \cudacpp{} \textit{can} be recycled for the brunt of amplitude calls at NLO. Now that an NLO multi-event interface exists, there is no need to develop these from scratch.

In \cref{algo:mgevloop,algo:paraevloop}, we give an overview of the program flow for a phase space point at NLO in the \mg{} codebase (left) and our ``naive'' branch-free algorithm (right). \Cref{algo:mgevloop} is highly branching, making it unsuited for SIMD/SIMT architectures. We forego this branching in \cref{algo:paraevloop} at the cost of superfluous calculations, allowing for parallelism over lines 6 through 10 (omitting the MC over loop evaluations for now). While this implies wasted compute, on heterogeneous systems we expect the loss to be outweighed by the speed-up from remaining on the device for longer. A smarter (albeit memory-intensive and latency-bound) algorithm might store momenta passing different cuts separately and evaluate asynchronously one type of amplitude in parallel once sufficiently many points pass. We defer this discussion until we have an implementation with SIMD/SIMT amplitudes running.

\subsection{Implementation status}
\label{sec:status}

Recently, development of event-level data-parallel NLO computations in \mg{} was officially started. The original work plan can briefly be summarised in four stages:
\begin{enumerate}
    \item Write a multi-event interface for fixed order NLO computations\footnote{The interface for fixed order computations (cross section integration) and unweighted event generation at NLO are disparate but largely identical in \mg{}. The former simply omits some complications of full event generation.} as per \cref{algo:paraevloop}.
    \item Make an initial data-parallel implementation by multithreading over events.
    \item Inject \cudacpp{}-generated tree-level amplitudes to utilise SIMD and SIMT hardware.
    \item Port functionality over to event generation.
\end{enumerate}
As of now, we have a working prototype at step 2, i.e. an event-parallel fixed order implementation using multithreading. This prototype is currently being validated, but appears to be data-safe for tested processes and yields cross sections in agreement with \mg{} to (1) numerical precision when run with the same seed and no vectorisation, and (2) statistical precision when run with vectorisation, due to a shift in random number generation calls.

While validation efforts are taking place and code refactoring and optimisations are considered, we are planning for the injection of \cudacpp{}-generated tree-level amplitudes into NLO executables. Although there are slight differences in how amplitudes are called, the interface is sufficiently similar to LO that only minor modifications need to be made on both sides to enable this. We expect to have a prototype also of this within the near future.

\section{Outlook}
\label{sec:outlook}

In lieu of the need for higher-precision theoretical HEP predictions and the success in porting hard LO event generation to data-parallel architectures, parallelising also NLO computations to GPUs and vectorised CPUs appears to be an obvious next step. We have illustrated that implementing hardware acceleration at NLO using the same methodology as used for LO event generation (i.e. focusing initially on porting scattering amplitudes to SIMD and SIMT architectures) is an appropriate measure, although the expected speed-up is less than what can be observed for large-multiplicity processes at LO. Furthermore, we are currently working on implementing this for tree-level amplitudes. Additional complications arise when considering e.g. loop amplitudes, which generally may require FP128 instructions, but moving tree-level amplitudes to SIMD/SIMT hardware alone should provide a significant speed-up. 

Furthermore, we have an extensive work plan for development of event-level parallelism for NLO event generation in the \mg{} software suite, which we are already tackling. So far, we have implemented a multi-event interface for fixed order NLO computations and tested the data safety of the algorithm by multithreading over phase space points. The next steps of development involve recycling data-parallel tree-level scattering amplitude routines from the \cudacpp{} plugin within the NLO codebase and porting these developments to event generation. We expect to have presentable results in the near future.

\section*{Acknowledgements}
We thank Stefano Frixione for useful discussion regarding FKS subtraction and the relevance of FKS sector parametrisation to SIMD/SIMT event-parallelism at NLO. We also thank Rikkert Frederix for extensive assistance in disentangling the \mg{} NLO codebase.

%
%

\bibliography{bibliography}

\end{document}